\documentclass[12pt]{article}
\usepackage{multirow}
\usepackage{graphicx}
\usepackage[cp1251]{inputenc}
\input{epsf}
\newcommand{\boxed}{\fbox{\rule[0.cm]{0.cm}{0.3cm}\ \ }}
\begin{document}
\begin{center}
{\Large\bf sp(4,R)-systematics of atomic nuclei. F-multiplets and
nuclear structure }\\
\bigskip\bigskip

 S.~Drenska\footnote{e-mail sdren@inrne.bas.bg}, M.~I.~Ivanov, N.~Minkov \\
\smallskip

{\em Institute of Nuclear Research and Nuclear Energy,Bulgarian
Academy of Sciences,
72 Tzarigrad Road, 1784 Sofia, Bulgaria} \\
\medskip

\title*{sp(4,R)-systematics of atomic nuclei. F-multiplets and
nuclear structure}

\author{{S.~Drenska}\inst{1}, \underline{{M.~I.~Ivanov}}\inst{1},
 \and {N.~Minkov}\inst{1}}
\end{center}
\bigskip\bigskip
\begin{abstract}
A systematics of the atomic nuclei in the  frame of the nucleon
number $A = Z + N$ and the proton-neutron difference $F = Z - N$
is considered. The classification scheme is provided by means of
the non-compact algebra $sp(4,R)$. In this scheme the nuclei are
ordered into isobaric multiplets, for which $A=\mbox{fix}$, as
well as in F-multiplets, for which $F=\mbox{fix}$. The dependence
of the mass excess $\Delta$, the first exited states $E_{2^+}$ and
the ratio $R_2=E_{4^+}/E_{2^+}$ on the nucleon number $A$ is
empirically investigated within the $F$-multiplets. Appropriate
filters are used to study the properties of the mass excess. Many
structural effects are observed. The mirror symmetry is clearly
indicated for the energy levels of the  nuclei with the same value
of $A$ and opposite $F$-values.
\end{abstract}
\newpage

\section{Introduction}
Usually the chart of the atomic nuclei is imaged  in the plane of
the proton number $Z$ and the neutron number $N$. It is naturally,
because the protons and neutrons are the particles from which the
nucleus  is composed and the levels of the shell structure are
given by their number. On the other hand there are important cases
when it is suitable to use the nucleon number $A = Z + N$ and $F =
Z - N$. For example: the Weizs\"{a}cker-Bethe mass formula
\cite{WBMF}; the description of the stability line
\cite{stability}, \cite{Nilsson}, \cite{Molnix} the isospin
symmetry, etc. Here, it is proposed to consider a systematics of
the atomic nuclei in the  framework of the atomic number $A=Z+N$
and  the proton-neutron difference $F=Z-N$, together with (and not
instead of) the systematics, based on $Z$ and $N$. This idea is
long known. First Ivan Selinov\cite{Selinov} made in 1948 a table
of nuclei  by using the coordinates $A$ and $\frac{1}{2}(Z - N)$.
This approach allows one to map the nuclei into the spaces of the
two irreducible infinite oscillator representations of the
non-compact algebra $sp(4,R)$. One can systematize the even-even
and odd-odd nuclei ($A$ - even) along the first one and the
even-odd and odd-even nuclei ($A$ - odd) along the order. The
proposed systematics is suitable for study of the nuclear mass
excess $\Delta$ and the half-life $T_{1/2}$. In particular, the
behaviour of $\Delta$ as a function of $F$ at $A=\mbox{fix}$ has
the known parabolic form in a very wide interval (up to $A =
260$). In the case of isobaric multiplets with even $A$, the mass
excess $\Delta$, considered as a function of $F$, exhibits a
staggering behaviour, corresponding to the alternation of the
even-even and odd-odd nuclei. For even $A$ isobaric multiplets
with $A \leq 208$ and for odd $A$ isobaric multiplets with $A \leq
209$ and $ 229 \leq A \leq 253$ both, the minimum of the mass
excess $\Delta$ and the maximum of the half-life $T_{1/2}$, are at
the same value of $F$. For the odd $211 \leq A \leq 227$ this rule
is not fulfilled, while for the even $A \geq 210$ and the odd $A
\geq 255$ the situation is ambiguous. The behaviour of $\Delta$ as
a function on $A$ for a given F-multiplet $(F = \mbox{fix})$  is
of a special interest. The corresponding curves $\Delta = \Delta
(A)|_{F=\mbox{\scriptsize fix}}$ are examined directly, as well as
with the help of their first and second discrete derivatives and
also through a specially constructed discrete function. All
considered curves show periodically repeating properties. All $Z$
and $N$ magic numbers giving the major shells are displayed by
distinct changes in the behaviour of the analyzed curves. Also, a
set of sub-magic numbers (giving sub-shells) as 6, 40, 64, etc is
well seen.  Noticeable changes in the behaviour of the curves are
observed at other values of $Z$ and $N$, such as 18, 60, 56, etc.
 They can be interpreted as  signs of possible
substructures. The common impression is that the curves $\Delta =
\Delta(A)_{F=\mbox{\scriptsize fix}}$ together with the
corresponding filters  contain a lot of information which needs to
be decoded and explained.

The dependence of the first exited states $E_{2^+}$ and the ratio
$R_2=E_{4^+}/E_{2^+}$ of even-even nuclei on the atomic number $A$
at $F=\mbox{fix}$ are empirically investigated. All major magic
and  doubly magic numbers are  clearly displayed in the
$F$-curves. Also a set of candidates for sub-magic numbers (giving
sub-shells), especially $Z=14,16,40$; $N=14,16,38,40$ is well
seen. An interesting behaviour for the nuclei with $N=88$:
$^{154}_{66}Dy_{88}$, $^{152}_{64}Gd_{88}$, $^{150}_{62}Sm_{88}$,
$^{148}_{60}Nd_{88}$, $^{146}_{58}Ce_{88}$, is observed.

 The  first four  nuclei of this
series are ``left'' neighbours in the corresponding $F$-curves of
the nuclei $^{156}_{66}Dy_{90}$, $^{154}_{64}Gd_{90}$
$^{152}_{62}Sm_{90}$, $^{150}_{60}Nd_{90}$, which are considered
as candidates for $X(5)$-nuclei (see \cite{Casten} and
\cite{DyGdSmND}).

A symmetry of the excitation levels of mirror nuclei with respect
to the inversion of the proton-neutron difference $F$ is clearly
observed: $E_{J^{\pi}}(A,F)=E_{J^{\pi}}(A,-F)$, where $J$ and
$\pi$ are the angular momentum and the parity, respectively, of
the states belonging to the same band \cite{lenziFe50}.

\section{$sp(4,R)$ - representations}

In the space of the nucleon number $A =Z+N$ and the proton-neutron
difference $F=Z-N\equiv 2T_0$ ($T_0$ is the third projection of
the isotopic spin) the nuclear chart splits into two parts: 1) $A$
and $F$ are even; 2) $A$ and $F$ are odd. This splitting makes
 it possible to map the nuclei into the spaces $H_+$ and
$H_{-}$ of the two irreducible infinite oscillator representations
of the non-compact algebra $sp(4,R)$ \cite{MPRA}. The nuclei with
even $A$ are mapped in $H_+$, while those with odd $A$ are
situated in $H_{-}$. This is illustrated schematically for the
case of $H_+$ in Table 1.
\begin{table}
\caption{\bf{Shematic structure of
the space $H_+$}}
\begin{center}
\begin{tabular}{|c|ccccccccccccc|}
\hline
A$\backslash$F&$\cdots$ & 10& 8& 6& 4& 2& 0& -2& -4& -6& -8& -10 &$\cdots$\\
\hline
0& & & & & & &$\boxed{   }$ & & & & & &\\
2& & & & & &$\boxed{   }$&$\boxed{   }$&$\boxed{   }$ & & & & &\\
4& & & & &$\boxed{   }$&$\boxed{   }$&$\boxed{   }$&$\boxed{   }$&$\boxed{   }$ & & & & \\
6& & & &$\boxed{   }$&$\boxed{   }$&$\boxed{   }$&$\boxed{   }$&
$\boxed{   }$&$\boxed{   }$&$\boxed{   }$ & & &\\
8& & &$\boxed{   }$&$\boxed{  }$&$\boxed{  }$ & $\boxed{   }$&
 $\boxed{   }$ & $\boxed{   }$&$\boxed{   }$&$\boxed{   }$&$\boxed{   }$ & & \\
$\vdots$&$\cdots$&$\cdots$&$\cdots$&$\cdots$&$\cdots$&$\cdots$&
$\cdots$&$\cdots$&$\cdots$&$\cdots$&$\cdots$&$\cdots$&$\cdots$\\
\hline
\end{tabular}
\end{center}
\end{table}

In this scheme the nucleon number $A$ is the first order Casimir
operator of the ``isobaric'' compact subgroup $u(2)\subset sp(4,R)$.
The nuclei are ordered in isobaric multiplets (isobars),
corresponding to the irreducible representations of $u(2)$ given by
the values of $A$ (the rows in Table 1). Till now there are
evidences for the existence of 294 isobars ($A=1,...,294$). On the
other hand, $F$ is interpreted as the first order Casimir operator
of the noncompact subgroup $u(1,1)\subset sp(4,R)$. The values of
$F$ give the oscillator representations of $u(1,1)$, according to
which the nuclei are ordered in F-multiplets (the columns in Table
1). Till now there are evidences for the existence of 70
$F$-multiplets ($F=8,...,1,0,-1,...,-61$).

\section{F-multiplets}

Let us consider the curves, giving the behaviour of the mass
excess $\Delta$ as a function on $A$ at $F = \mbox{fix}$ or in
other words the behaviour of $\Delta$ within the F-multiplets. We
shall refer to the corresponding curves as $F$-$\Delta$-curves. In
Fig.~1 and Fig.~2 examples of F-$\Delta$-curves at $F=0, -4, -23 ,
-44$ are given. (All experimental data on the mass excess
$\Delta$, in MeV, are taken from \cite{ensdf}.) These examples
illustrate the main characteristics of the $F$-$\Delta$ -curves:

\begin{figure}[t]
\centerline{{\epsfxsize=8.cm\epsfbox{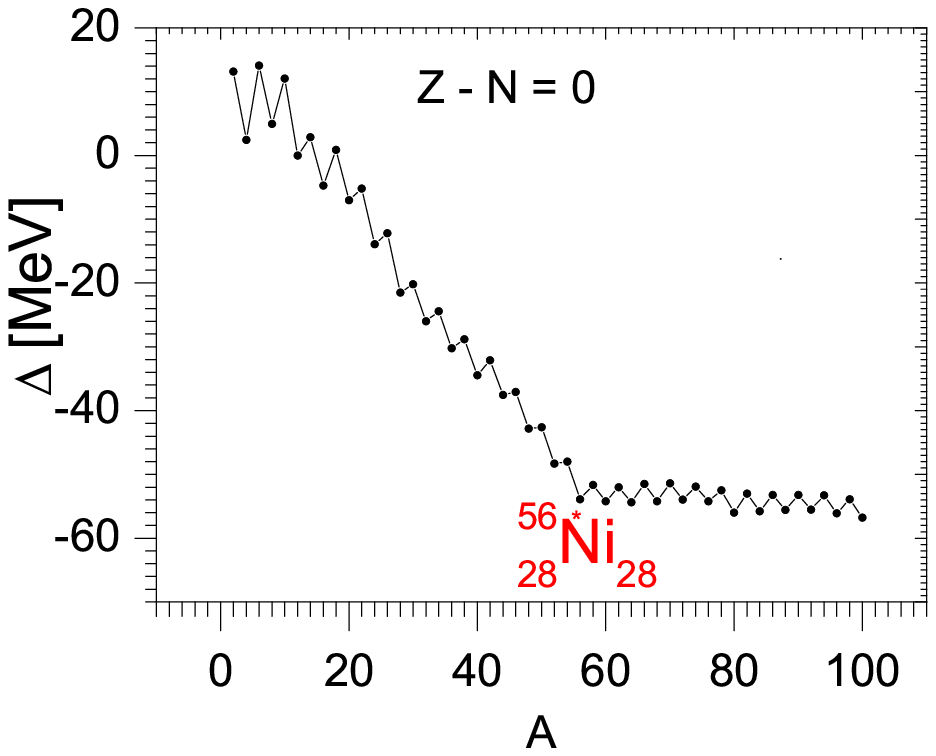}}\ \
{\epsfxsize=8.cm\epsfbox{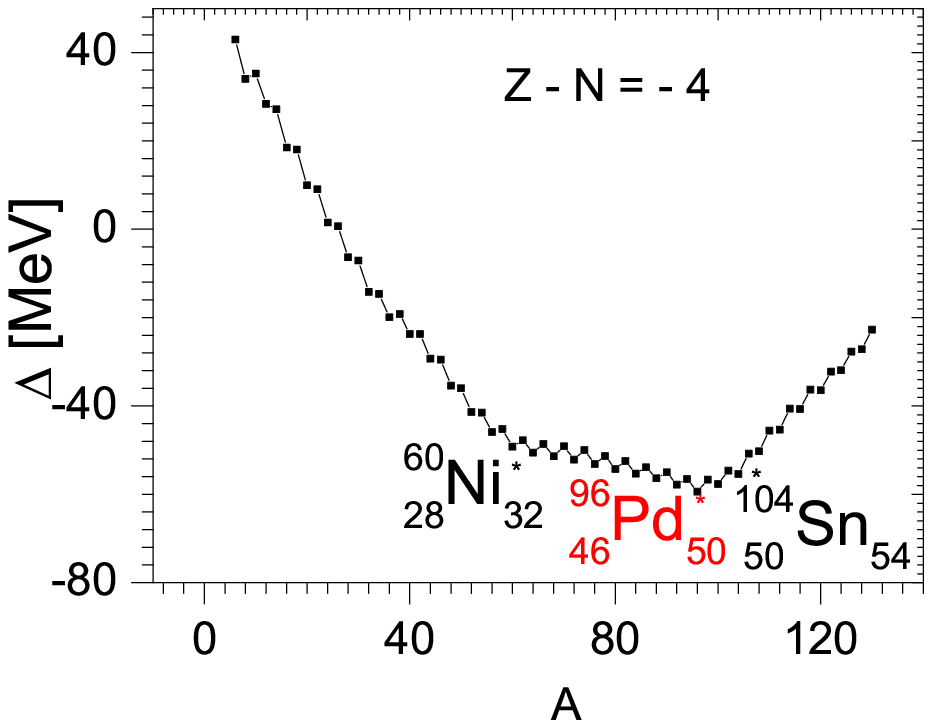}}} \caption{The mass excess
$\Delta$ as a function of $A$ for $F=0$
 (left) and $F=-4$ (right).} \label{fig:01}
\end{figure}

\begin{figure}[t]
\centerline{{\epsfxsize=8.cm\epsfbox{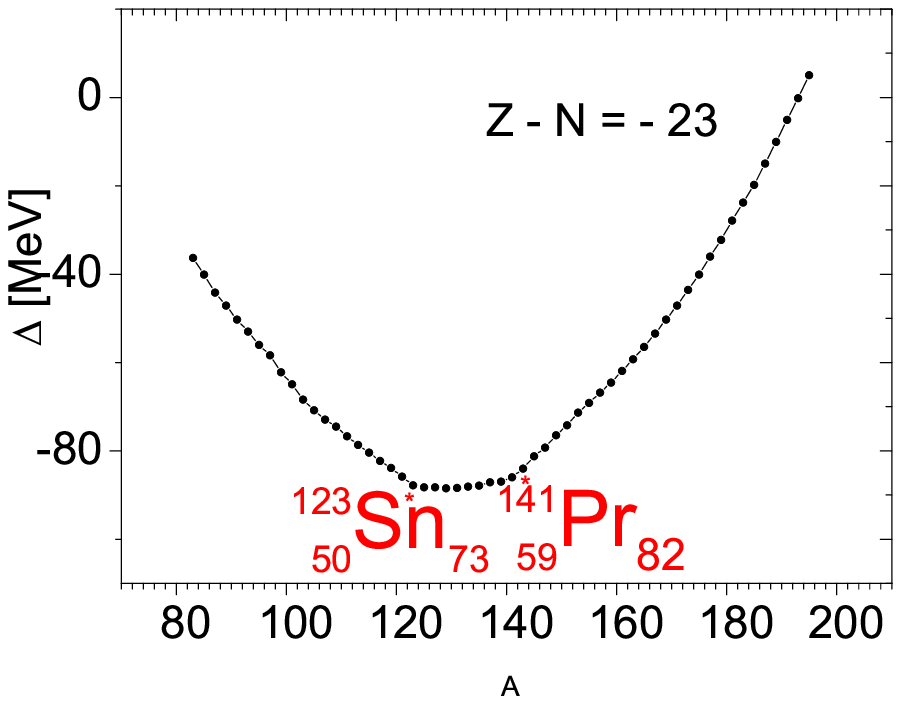}}\ \
{\epsfxsize=8.cm\epsfbox{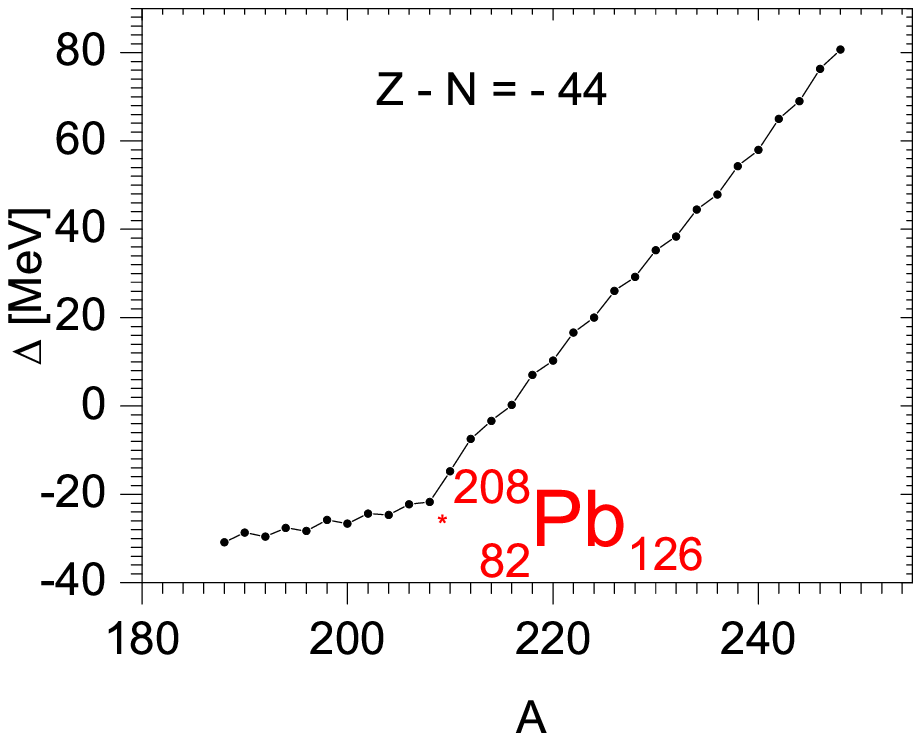}}} \caption{$\Delta(A)$ for
$F=-23$ (left) and $F=-44$ (right).} \label{fig:02}
\end{figure}

\begin{itemize}
\item The values of $A$ for which  the behaviour of a given F-$\Delta$- curve
changes  considerably correspond to major magic $Z$-numbers or/and
 to major magic $N$-numbers. These are: $^{56}_{28}Ni_{28}$ at
$F=0$; $^{60}_{28}Ni_{32}$ and $^{96}_{46}Pd_{50}$  at $F=-4$;
$^{123}_{50}Sn_{73}$ and $^{141}_{59}Pr_{82}$ at $F=-23$;
$^{208}_{82}Pb_{126}$ at $F=-44$.
\item The ``odd'' curves ($F$ is odd) have  relatively
smoother behaviour while the ``even'' curves ($F$ is even) have
 well seen sectors of a ``staggering'' and a ``coupling'' behaviour.
It is known that analogical observation takes place in the case of
isobaric multiplets. Note that this difference corresponds to the
splitting of the nuclear chart into two subspaces, $H_+$ and
$H_-$.
\end{itemize}

 The staggering behaviour corresponds to alternating change
up/down in the discrete-function value with the changing discrete
values of the argument. In the case of even $F$ the staggering
corresponds to a splitting of the curve into two smoother curves,
``even-even'' and ``odd-odd'' ones. The coupling behaviour
corresponds to alternating change short/long of the distance
between two neighbouring points of a given sector of the discrete
curve. In the case of even $F$ the set of nuclei corresponding to
this sector splits into couples each of them containing one
even-even and one odd-odd nuclei.
\begin{itemize}
\item
As a rule, when the curve has a clear minimum, this minimum
corresponds to a nucleus with major magic $Z$ or/and magic $N$
($^{96}_{46}Pd_{50}$  at $F=-4$). From the other hand when $F$ is
even the minimum plays a role of a \textbf{reversal point} in
which the order of the nuclei in the couple changes from
(even-even,odd-odd) to (odd-odd,even-even) (There are exceptions,
especially when the minimum is not distinctive.)
\item
 In the case of the ``even'' $F$-$\Delta$-curves the direction
of the segments of the curve between the ``coupled points''
clearly changes at the values of $A$ corresponding to nuclei with
major magic $Z$ or/and magic $N$ numbers (e.g.
$^{104}_{50}Sn_{54}$ at $F=-4$). However, such changes take place
also at other values of $A$, pointing out the need of further
study of the $F$-$\Delta$-curves.
\end{itemize}
 For more detailed investigations of the
$F$-$\Delta$-curves we introduce the following filters:

  1. The first discrete derivative:
\begin{eqnarray}
    D_1[\Delta(A)] =   1/2[\Delta(A) - \Delta(A+2)]. \\  \nonumber
\end{eqnarray}
 The staggering of the first discrete derivative
corresponds to coupling and/or staggering behaviour of a given
$F$-$\Delta$-curve. When the staggering curve $D_1[\Delta(A)]$
does not cross the axis $\Delta=0$, the curve has only a coupling
behaviour. When the staggering curve $D_1[\Delta(A)]$ crosses the
axis, the $F$-$\Delta$-curve has certainly a staggering, which may
be combined  or not combined with the coupling behaviour. Compare
Fig.~3, left with Fig.~3, right.

\begin{figure}[t]
\centerline{{\epsfxsize=8.cm\epsfbox{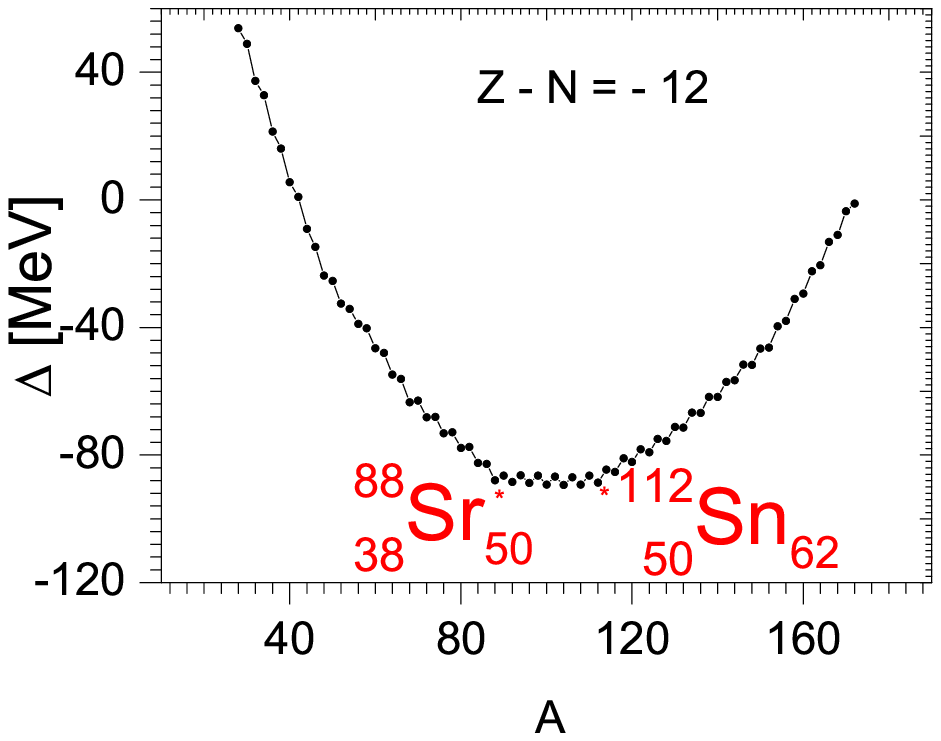}}\ \
{\epsfxsize=8.cm\epsfbox{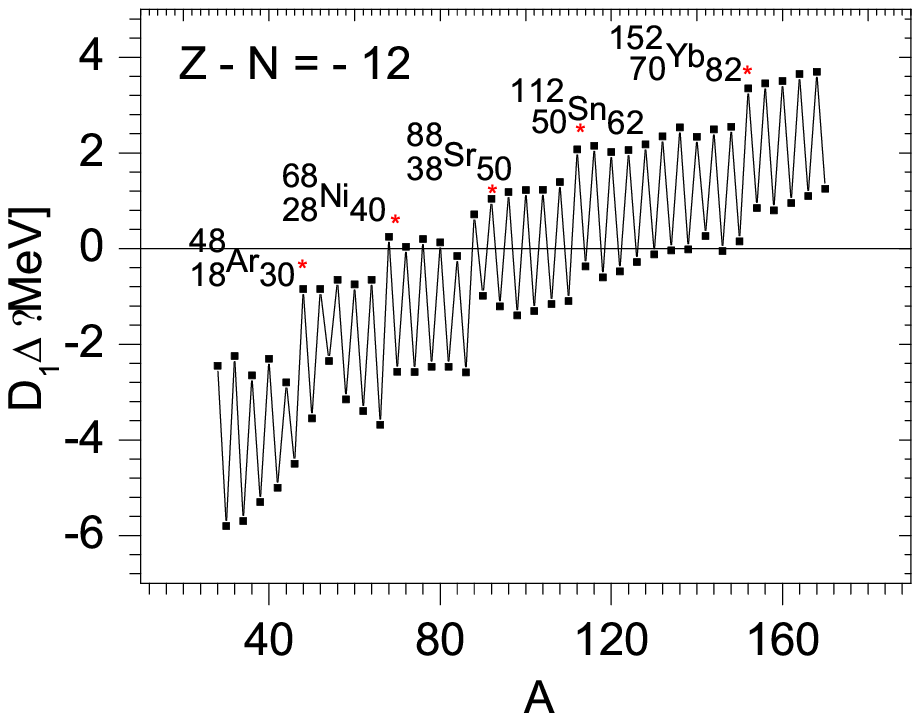}}} \caption{$\Delta(A)$ for
$F=-12$ (left); $D_1\Delta(A)$ for $F=-12$ (right).}
\label{fig:03}
\end{figure}

2. The modulus $|D_1[\Delta(A)]|$ of the first discrete
derivative:

  It clearly indicates the sector, where the staggering behaviour
dominates ($A=88,...,112$ at $F=-12$) and sectors with only
coupling behaviour given by $88 \leq A$ and $A \geq 112$ at
$F=-12$ (compare Fig.~3 (left) with Fig.~4 (left)). The function
$|D_1[\Delta(A)]|$ indicates also \textbf{reversal points}, where
the coupling is changed from ``left to right'' to ``right to
left'' (at the reversal point the phase of the staggering
changes).

3. The second discrete derivative:
\begin{eqnarray}
 D_2[\Delta(A)]=1/4[ \Delta(A+2) - 2\Delta(A)
 + \Delta(A-2)].
\end{eqnarray}
 It indicates both the coupling and the staggering
effects, but can not distinguish them, when they take place
simultaneously. From the other hand this filter is strongly
sensitive to the deviations. As a rule  these deviations are
displayed at the values of $A$, which correspond to the magic or
sub-magic numbers (see Fig.~4, right).
\begin{figure}[t]
\centerline{{\epsfxsize=8.cm\epsfbox{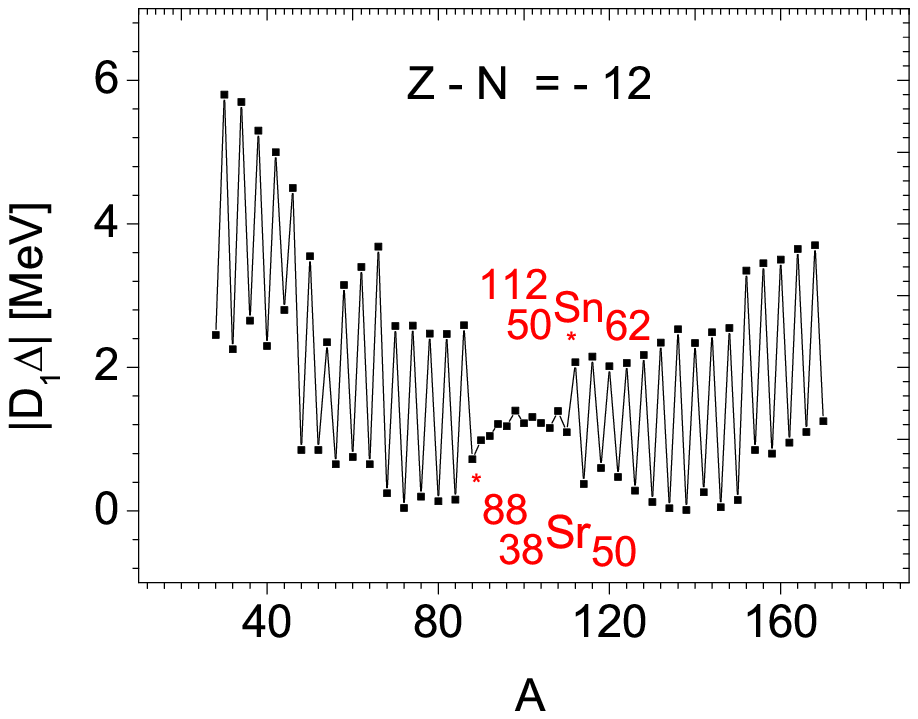}}\ \
{\epsfxsize=8.cm\epsfbox{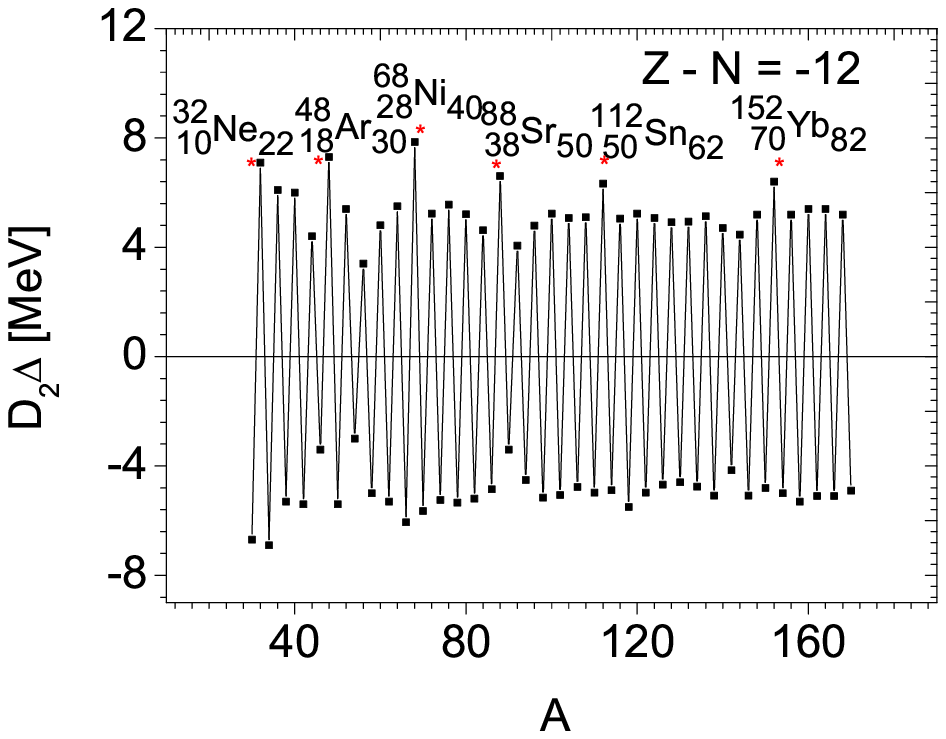}}} \caption{$|D_1[\Delta(A)]|$
for $F=-12$ (left); $D_2\Delta(A)$ for $F=-12$ (right).}
\label{fig:04}
\end{figure}

4. D-function:

Let us introduce the function:
\begin{eqnarray}
    D[\Delta(A)] =  | 1/2([\Delta(A) - \Delta(A+2)]
   - [\Delta(A-4) - \Delta(A-2)])|,
\end{eqnarray}
for $A = A_{0} + 4$, $A_{0} + 8$, ... ,$A_{min}  - 4$, \\
\begin{eqnarray}
    D[\Delta(A)]  = | 1/2([\Delta(A) - \Delta(A-2)]
     -[\Delta(A+4) - \Delta(A+2)])|,
\end{eqnarray}
for $A =A_{min}, A_{min} + 4$, $A_{min} + 8$, ..., $A_f  - 4$, \\
where $A_0$ and $A_f$  are the values of $A$ corresponding to the
left point of the first couple of the $F$- $\Delta$-curve and the
right point of the last couple of the $F$-$\Delta$ - curve
respectively; $A_{min}$ is defined by $\Delta(A_{min}) = min
\Delta(A)$.

\begin{figure}[t]
\centerline{{\epsfxsize=8.cm\epsfbox{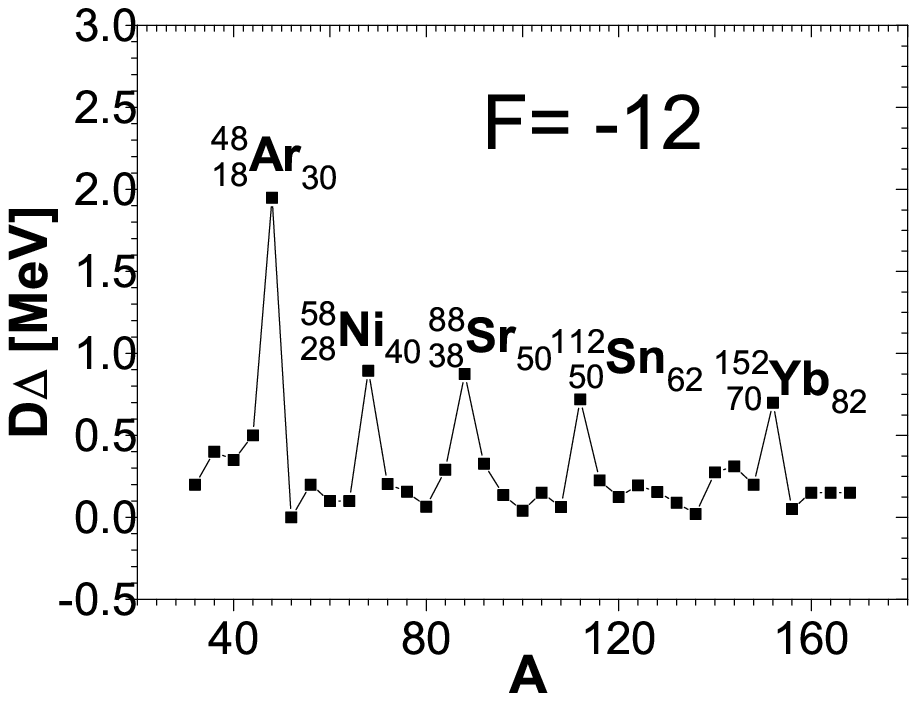}}\ \
{\epsfxsize=8.cm\epsfbox{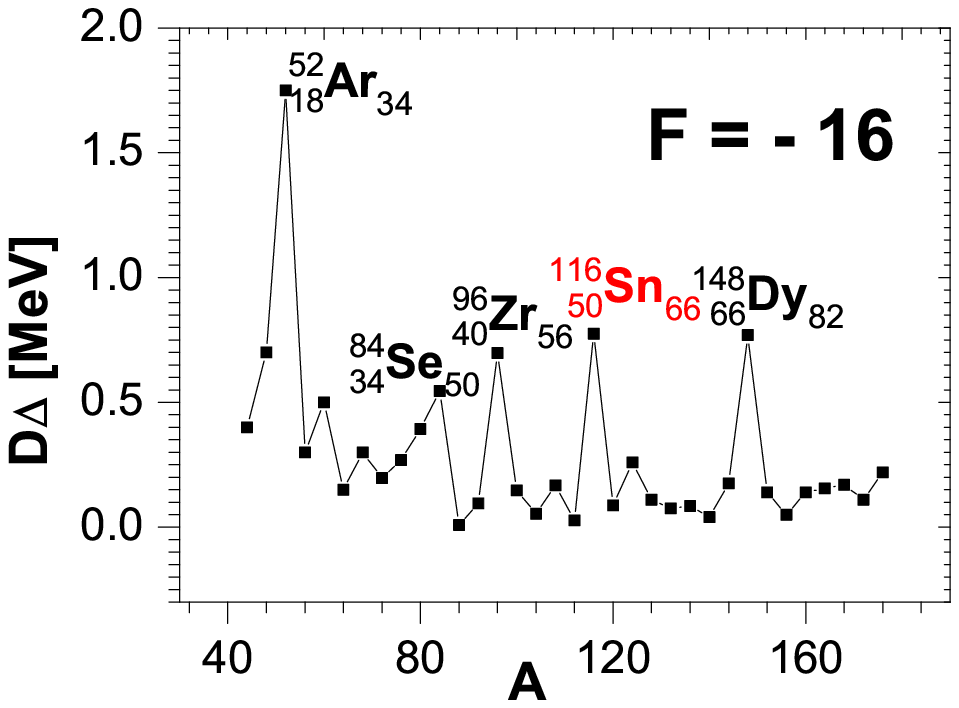}}} \caption{$D$-function for
$F=-12$ (left); $D$-function for $F=-16$
 (right).} \label{fig:05}
\end{figure}

 The D-function indicates the changes of the direction of the segments
between ``coupled points'' (see Fig.~5,left and right).

All filters are used  here only for even multiplets.

The $F$-$\Delta$- curves and the corresponding filters contain a
lot of information:
\begin{itemize}
\item Clear indications for the existence of the shell structure
of the nuclei are seen. All $Z$ and $N$ magic numbers giving the
major shells are displayed by distinct changes in the behaviour of
the $F$-$\Delta$-curves and of the introduced filters
\item A set of sub-magic numbers (giving sub-shells) as 40, 64, etc.
is well seen.
\item Noticeable changes in the behaviour of the filters under
consideration are observed at other values of $Z$ and $N$ such as
18, 56, 60, etc. These effects need to be explained.
\end{itemize}
\newpage
\section {$F$-$E_{2_+}$-curves and $F$-$R_2$-curves}
In the case of the even-even nuclei a lot of information is
contained also in the curves which give the dependence of
$E_{2_+}$ and $R_2=E_{4_+}/E_{2_+}$  on $A$ at $F=\mbox{fix}$.
($E_{4_+}$ and $E_{2_+}$ are from the ground band states.) We
shall call these curves $F$-$E_{2_+}$ - curves and
$F$-$R_2$-curves respectively. The bundle of all known
$F$-$E_{2_+}$ - curves are displayed in Fig.~6. This picture is an
other example, which shows the advantage of the systematics of the
nuclei in $F$-multiplets:
\begin{itemize}
\item  The major magic and  doubly magic numbers are well seen  as
strong peaks of the $F$-$E_{2_+}$-curves and deep minimums of the
$F$-$R_{2}$-curves. Some sub-magic nuclei as $^{96}_{40}Zr_{56}$ are
also clearly displayed (see Fig.~6).
\item The even-even nuclei are grouped in shell multiplets.
It is especially well seen for the major shells: $(28,28|50,50)$,
$(28,50|50,82)$, $(50,50|82,82)$,; $(50,82|82,126)$,
$(82,126|126,?)$. (We denote the given major shell with
$(MZ_i,MN_j|MZ_k,MN_l)$, where $M Z_i,MN_j$ and $M Z_k,MN_l$ are
two ``neighbour'' double magic numbers defining uniquely the shell
$(MZ_i<MZ_k$, $MN_j<MN_l)$.
\end{itemize}

\begin{figure}[t]
\epsfxsize=14.cm   
\centerline{\epsfbox{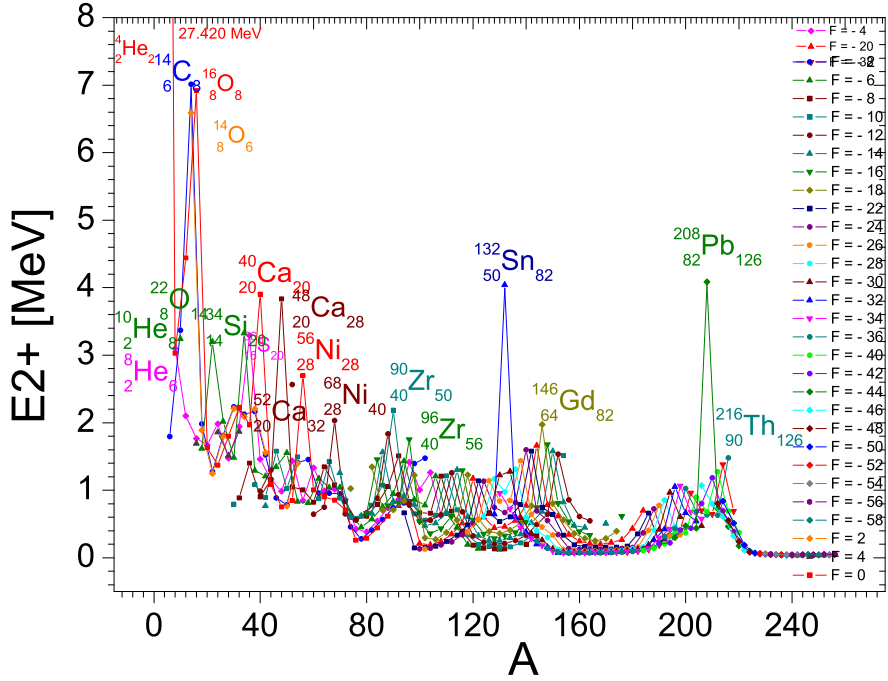}} \caption{The behaviour of all
$F$-$E_{2_+}$-curves.
 The experimental data are from
\cite{ensdf}.} \label{fig:06}
\end{figure}

An analogical analysis can be done for the bundle of all
$F$-$R_2$-curves.

More detailed investigation can be provided if we consider the
segments of the $F$-$E_{2_+}$-bundle and the $F$-$R_2$-bundle in a
given major shell \cite{M1}. The picture of the segments of the
$F$-$E_{2_+}$-bundle and the $F$-$R_2$-bundle in the framework of
a given major shell presents a detached configuration. Usually,
$F$-$E_{2_+}$-curves in a such segment decrease monotonously from
the left side (inhabited by magic nuclei) to the bottom and after
that go up monotonously to the right side (inhabited also by magic
nuclei). In the case of heavy nuclei the ``bottom'' is flat in a
very long interval. As to the $F$-$R_2$-curves, they increase
monotonously from the left side (inhabited by magic nuclei) to the
``roof'' and after that go down monotonously to the right side
(inhabited by magic nuclei also). In the case of heavy nuclei the
``roof'' is flat in a very long interval. In many large areas the
curves do not intersect to each other, but the external curve
embraces the internal one.

\begin{figure}[t]
\epsfxsize=14.cm   
\centerline{\epsfbox{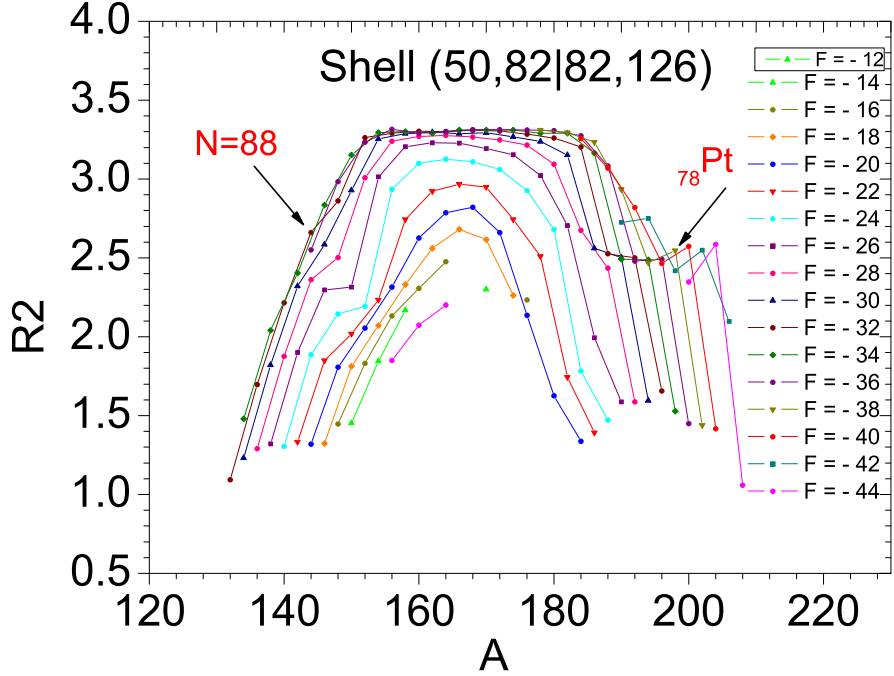}} \caption{The behaviour in all
$F$-multiplets passing through the shell $(50,82|82,126)$.}
\label{fig:07}
\end{figure}

But, what we are looking for are the exceptions (the deviations)
from this ``right'' behaviour. Let us point out several examples
for deviations observed in different shells:

\begin{figure}[t]
\centerline{{\epsfxsize=8.cm\epsfbox{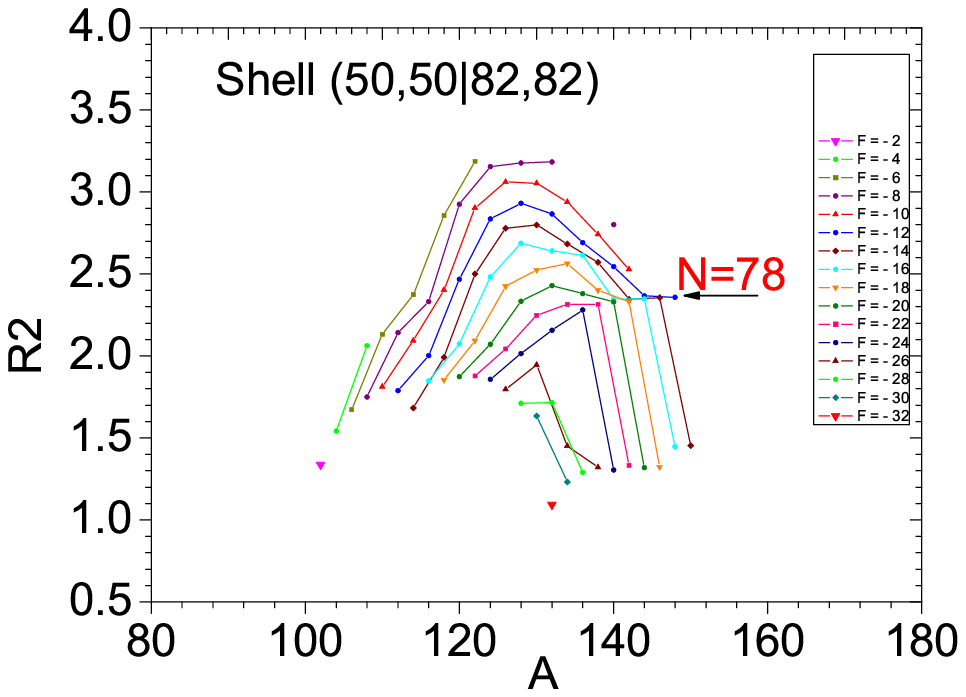}}\ \
{\epsfxsize=8.cm\epsfbox{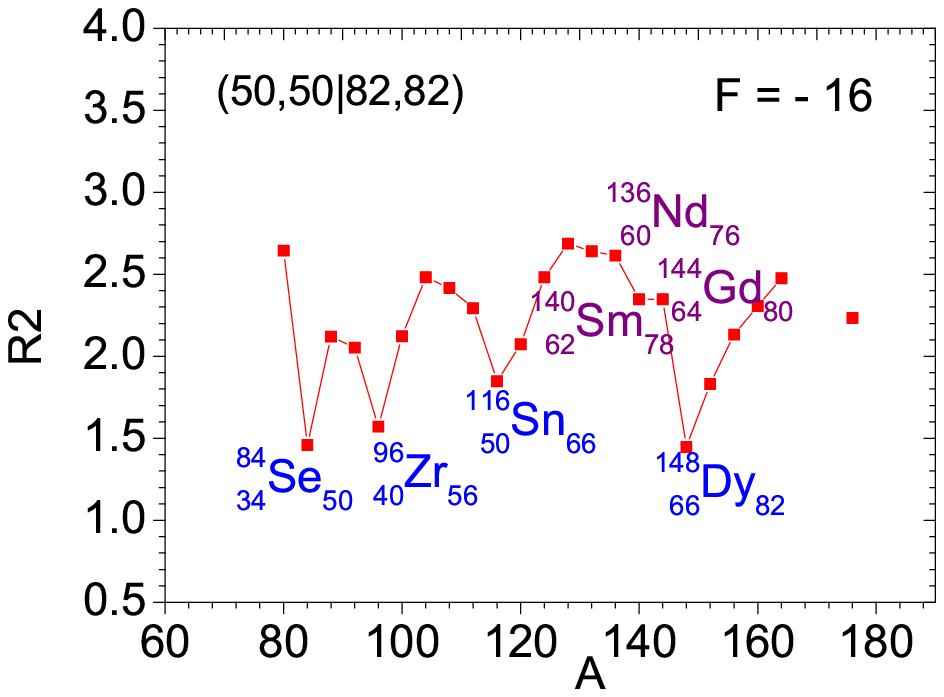}}} \caption{The bundle
$R_2$-curves for all $F$-multiplets passing in the shell
$(50,50|82,82)$ (left), the $R_2$-curve for $F=-16$ (right).}
\label{fig:08}
\end{figure}

In the shell $(28,28|50,50)$ there are  deviations  at
$F=0,-2,-4,-6,-8$ and the nuclei: \\
 $^{70}_{30}Zn_{40}$. $^{64}_{32}Ge_{32}$, $^{66}_{32}Ge_{34}$,
 $^{68}_{32}Ge_{36}$,$^{70}_{32}Ge_{38}$,$^{72}_{32}Ge_{40}$ \\
 $^{68}_{34}Se_{34}$,$^{70}_{34}Se_{36}$,
$^{72}_{34}Se_{38}$ \\ $^{72}_{36}Kr_{36}$,$^{74}_{36}Kr_{38}$.

\begin{figure}[t]
\centerline{{\epsfxsize=8.cm\epsfbox{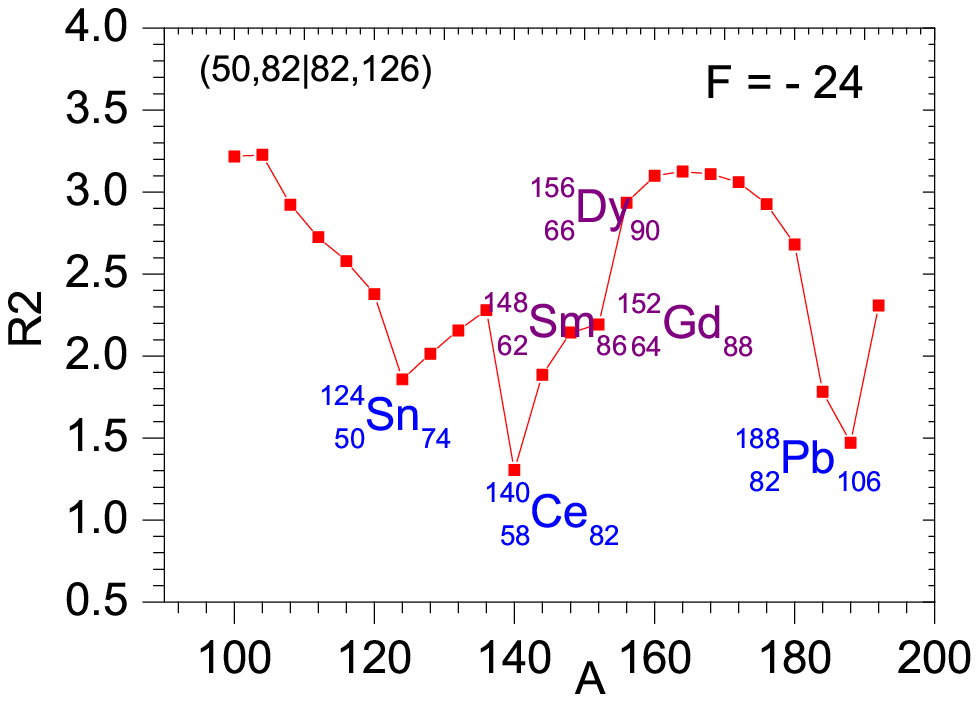}}\ \
{\epsfxsize=8.cm\epsfbox{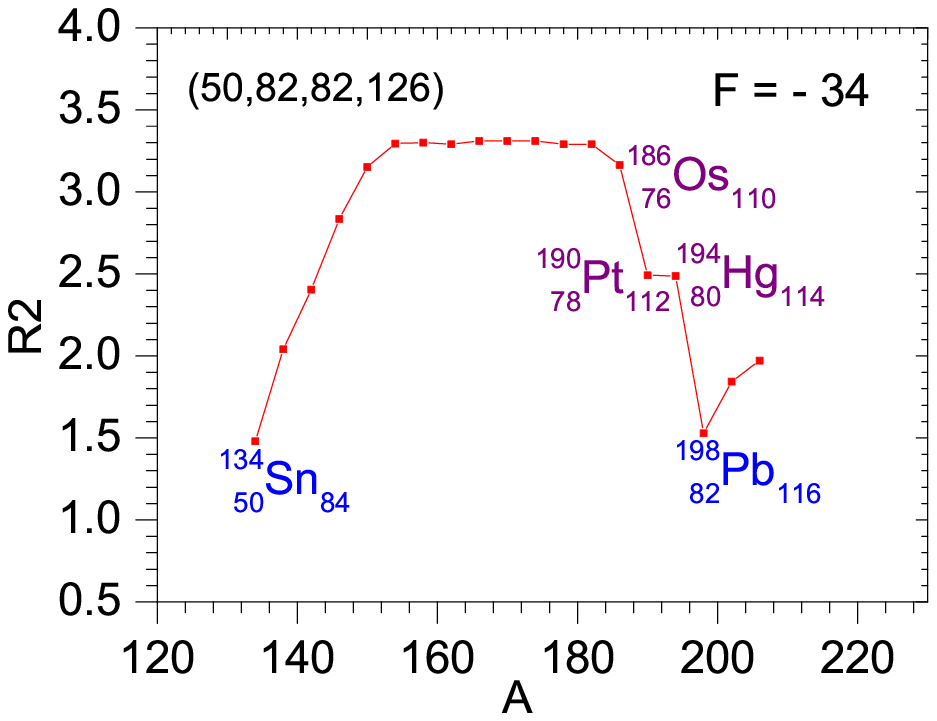}}} \caption{The behaviour of
$R_2$-curves for the $F=-24$ (left) and for $F=-34$ (right).}
\label{fig:09}
\end{figure}

In the shell $(28,50|50,82)$ the strong deviations are observed at
$F=-12,-14,-16,-18$ and the nuclei: \\
 $^{96}_{38}Sr_{58}$ \\
 $^{92}_{40}Zr_{52}$,
$^{94}_{40}Zr_{54}$, $^{96}_{40}Zr_{56}$,  $^{98}_{40}Zr_{58}$,
\\ $^{96}_{42}Mo_{54}$, $^{98}_{42}Mo_{56}$.

A possible explanation of these deviations is an existence of a
sub-shell. In the shell $(50,50|82,82)$ there are  deviations at
$F=-12,-14,-16,-18,-20$ and the nuclei:\\
$^{144}_{66}Dy_{78}$, $^{142}_{64}Gd_{78}$, $^{140}_{62}Sm_{78}$,
$^{138}_{60}Nd_{78}$, $^{136}_{58}Ce_{78}$  $N=78$-series, (see
Fig.~8, left and right).

In the shell $(50,82|82,126)$ there are deviations at \\
$F=-22,-24,-26,-28,-30$ and the nuclei: $^{154}_{66}Dy_{88}$,
$^{152}_{64}Gd_{88}$, $^{150}_{62}Sm_{88}$, $^{148}_{60}Nd_{88}$,
$^{146}_{58}Ce_{88}$ - $N=88$-series (see Fig~7 and Fig~9, left).
\newpage
We remark that the first four nuclei of this series are ``left''
neighbours in the corresponding  $F$-$E_{2_+}$-curves  and
$F$-$R_2$-curves of the nuclei $^{156}_{66}Dy_{90}$,
$^{154}_{64}Gd_{90}$, $^{152}_{62}Sm_{90}$, $^{150}_{60}Nd_{90}$,
which are considered as $X(5)$- nuclei \cite{Casten} (see Fig.~9,
left).

There are also well expressed deviations in this shell at \\
$F=-28, -30,-32,-34,-36,-38,-40,-42$ and the nuclei:
 $^{184}_{78}Pt_{106}$,  $^{186}_{78}Pt_{108}$,
$^{188}_{78}Pt_{110}$, $^{190}_{78}Pt_{112}$,
 $^{192}_{78}Pt_{114}$,  $^{194}_{78}Pt_{116}$,
$^{196}_{78}Pt_{118}$,  $^{198}_{78}Pt_{120}$ $Pt$-series (see
Fig.~7 and Fig.~9, right).

Additional information can be extracted from $F$-$E_{2_+}$-curves
and $F$-$R_2$-curves, each of them presented entirely (see, for
example Fig.~8, right and Fig.~9, left and right). On these
pictures all magic and doubly magic numbers are displayed clearly.
Also, a set of candidates for sub-magic numbers (giving
sub-shells) is well seen, especially $Z=14,16,40$ and
$N=14,16,38,40$.
\newpage
\section{F-multiplets and mirror nuclei}

\begin{figure}[t]
\centerline{{\epsfxsize=8.cm\epsfbox{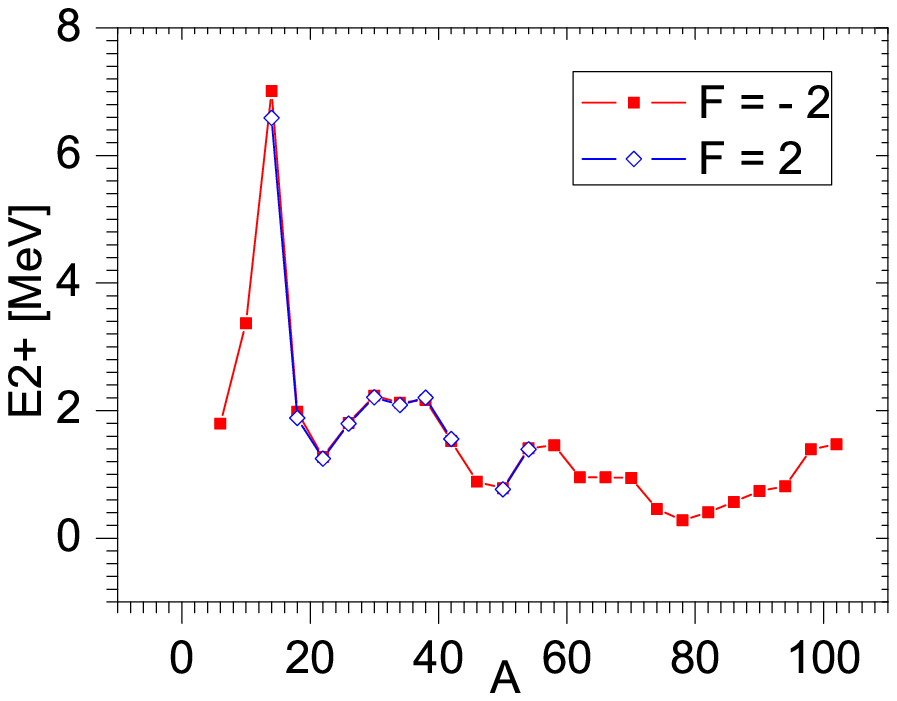}}\ \
{\epsfxsize=8.cm\epsfbox{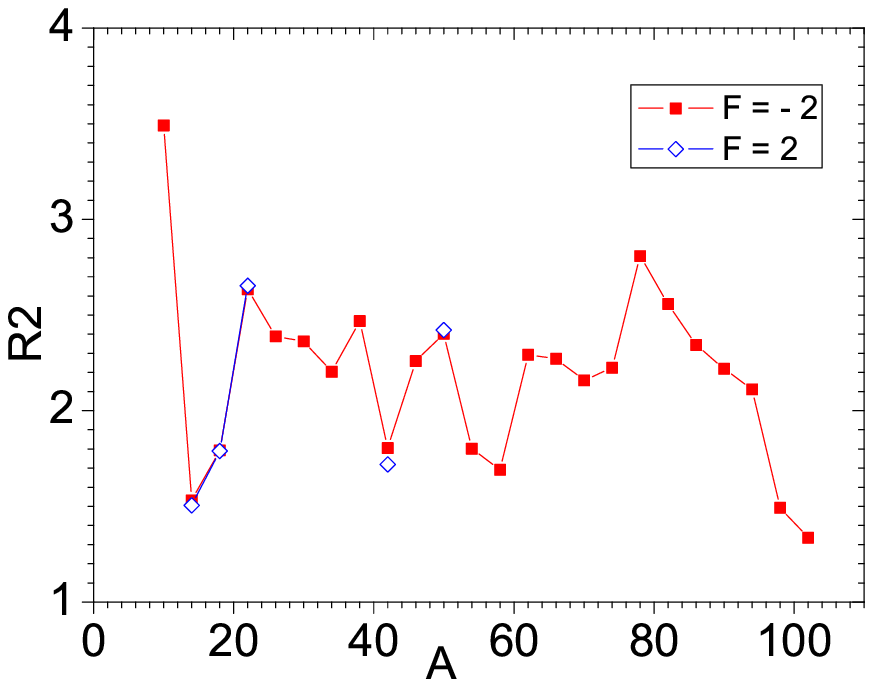}}} \caption{$E_{2^+}$ energy
states (left) and $R_2$-curves (right) for all mirror nuclei in
$F=\pm2$-multiplets.} \label{fig:10}
\end{figure}

The F-multiplets are relevant also to study the properties of the
mirror nuclei.  Mirror nuclei are two nuclei with the same nucleon
number $A$ but interchanged proton number $Z$  and neutron number
$N$. They can be observed in the $F$-multiplets with $F=\pm
8,7,6,5,4,3,2,1$. A  symmetry is observed between the energy
levels of the mirror nuclei: $E_{J^{\pi}}(A,F)=E_{J^{\pi}}(A,-F)$,
where $J$ is the angular momentum and $\pi$ is the parity of the
respective excited states of the same band. We examined this rule
for all data in available which are at $F=\pm1,...,\pm7$. The
ground states of all mirror nuclei are at the same values of
$J^{\pi}$. There are only three exceptions.These are the following
ground states: $^{16}_{9}F_{7} (0^-)$ and $^{16}_{7}N_{9} (2^-)$
at $F=\pm2$; $^{22}_{13}Al_{9} (3^+)$ and $^{22}_{9}F_{13} (4^+)$
at $F=\pm4$; $^{25}_{15}P_{10} (1/2^+)$ and $^{25}_{10}Ne_{15}
(3/2^+)$ at $F=\pm5$.
\newpage
\begin{figure}[t]
\centerline{{\epsfxsize=7.cm\epsfbox{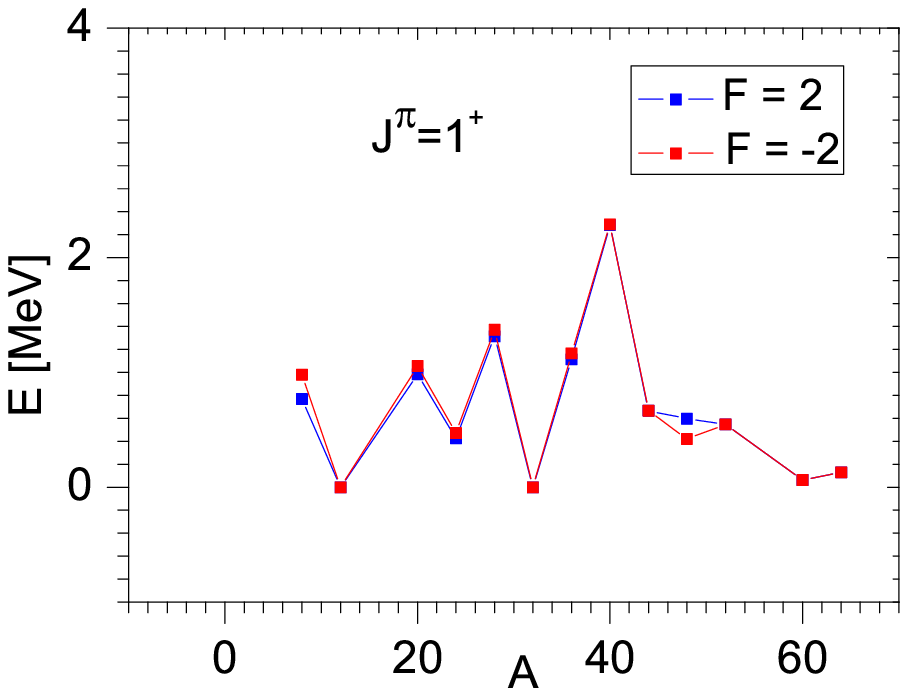}}\ \
{\epsfxsize=7.cm\epsfbox{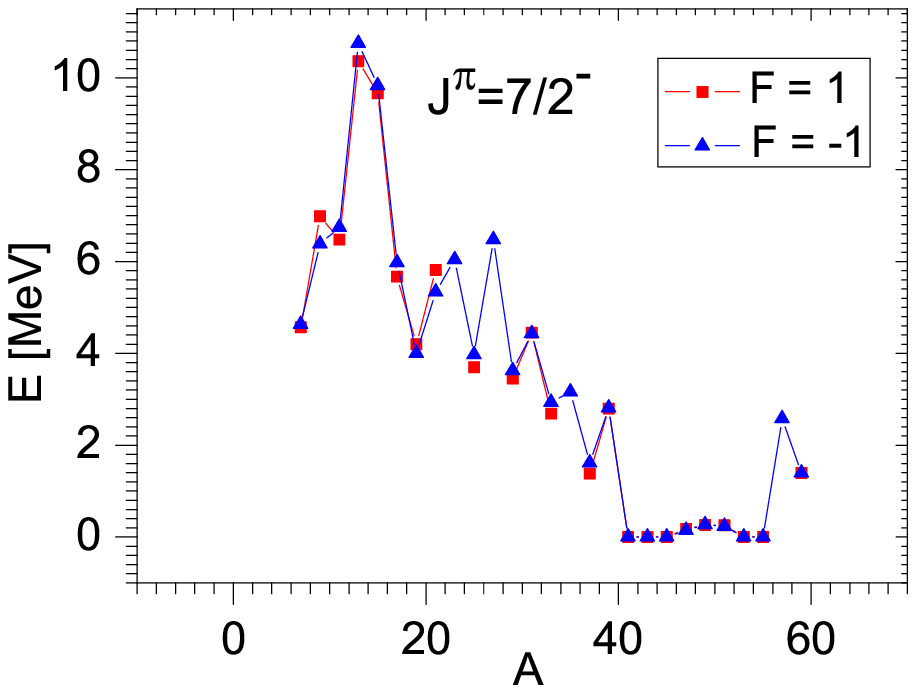}}} \caption{$E_{1^+}$ energy
states  for all the mirror nuclei in $F=\pm 2$ -multiplets (left);
$E_{7/2^-}$ energy states for all the mirror nuclei in $F=\pm 1$
(right).} \label{fig:11}
\end{figure}
The considered symmetry takes place for the data at $F=\pm2, \pm
4$ and $J^\pi= +2^+, 4^+,...,10^+$ of the ground band. The
$F$-$E_{2^+}$ and $F$-$R_2$-curves at $F=\pm2$ are given in
Fig.~10. Another example is given in Fig.~11, left for
$J^\pi={1^+}$ at $F=\pm2$. The  mirror energy symmetry is observed
also for many states of the even-odd and odd-even nuclei belonging
to the multiplets given by $F=\pm1,\pm3$. See for instance
Fig.~11, right, for the case $J^\pi=7/2^-$ at $F=\pm1$. The mirror
energy symmetry is suitable for the further investigation of the
proton-rich nuclei.
\newpage
\noindent {\Large \bf Acknowledgments} The authors are grateful to
the seminar  ``Theory of the atomic nuclei'' of JINR Dubna for
valuable discussion. This work is supported  by the Bulgarian
Scientific Fund under contract F-1502/05.


\begin{thebibliography}{xx}

\bibitem{WBMF} C. F. Weizs\"{a}cker, Z. Physik {\bf 96}, 431 (1935).
\bibitem{stability} T. Kodama, Prog. Theor. Phys. {\bf 45}, 1112 (1971).
\bibitem{Nilsson} S. G. Nilsson and I. Ragnarsson, ``Shapes and
shells in nuclear structure''. Cambridge University Press, (1995).
\bibitem{Molnix} P. M\"{o}ller, and J. R. Nix, W. D. Myers and W. J.
Swiatecki, Atomic Data and Nuclear Data Tables {\bf 59}, 185
(1995).
\bibitem{Selinov} I. P. Selinov, ``The structure and systematics
of atomic nuclei''. Nauka, Moscow, 1990.
\bibitem{Casten}R. F. Casten and E. A. McCutchan, J. Phys. G:Nucl. Part. Phys.
{\bf 34}, R285 (2007).
\bibitem{DyGdSmND} P. Petkov et al., Phys. Rev.C {\bf 68}, 034328 (2003).
\bibitem{lenziFe50} S. M. Lenzi et al., Phys.Rev. Lett. {\bf
87}, 122501 (2001).
\bibitem{MPRA} A. I. Georgieva, M. I. Ivanov, P. P. Raychev, and
R. P. Roussev, Int. J. Theor. Phys., {\bf 25}, 1181 (1986).
\bibitem{ensdf} http://www.nndc.bnl.gov/ensdf/.
\bibitem{M1}A. I. Georgieva, M. I. Ivanov, P. P. Raychev, and
R. P. Roussev, Int. J. Theor. Phys., {\bf 28}, 769 (1989).
\end{thebibliography}
\end{document}